\documentclass{emulateapj}

\newcommand{\citepeg}[1]{\citep[{e.g.,}][]{#1}}

\shorttitle{Radio Transient in Cygnus A}
\shortauthors{Perley et al.}

\begin{document}

\title{Discovery of a Luminous Radio Transient 460 pc from the Central \\ Supermassive Black Hole in Cygnus A}

\def\ljmu{1}
\def\dark{2}
\def\nrao{3}
\def\cav{4}
\def\mail{*}

\author{Daniel~A.~Perley\altaffilmark{\ljmu,\dark,\mail},
Richard~A.~Perley\altaffilmark{\nrao}, 
Vivek~Dhawan\altaffilmark{\nrao}, and
Christopher~L.~Carilli\altaffilmark{\nrao,\cav}}

\altaffiltext{\ljmu}{Astrophysics Research Institute, Liverpool John Moores University, IC2, Liverpool Science Park, 146 Brownlow Hill, Liverpool L3 5RF, UK}
\altaffiltext{\dark}{Dark Cosmology Centre, Niels Bohr Institute, University of Copenhagen, Juliane Maries Vej 30, 2100 K{\o}benhavn {\O}, Denmark}
\altaffiltext{\nrao}{National Radio Astronomy Observatory, P.O. Box O, Socorro, NM, 87801}
\altaffiltext{\cav}{Cavendish Astrophysics Group, Cambridge CB3 0HE, UK}
\altaffiltext{\mail}{e-mail: d.a.perley@ljmu.ac.uk }
\slugcomment{Accepted to ApJ 2017-05-08}

\begin{abstract}
We report the appearance of a new radio source at a projected offset of 460 pc from the nucleus of Cygnus A.  The flux density of the source (which we designate Cygnus A-2) rose from an upper limit of $<$0.5 mJy in 1989 to 4 mJy in 2016 ($\nu$=8.5 GHz), but is currently not varying by more than a few percent per year.  The radio luminosity of the source is comparable to the most luminous known supernovae, it is compact in VLBA observations down to a scale of 4 pc, and it is coincident with a near-infrared point source seen in pre-existing adaptive optics and HST observations.  The most likely interpretation of this source is that it represents a secondary supermassive black hole in a close orbit around the Cygnus A primary, although an exotic supernova model cannot be ruled out.  The gravitational influence of a secondary SMBH at this location may have played an important role in triggering the rapid accretion that has powered the Cygnus A radio jet over the past $10^7$ years.
\end{abstract}

\keywords{galaxies: individual: Cygnus A --- galaxies: nuclei --- galaxies: active --- quasars: supermassive black holes --- supernovae}

\section{Introduction}
\label{sec:intro}

The past decade has witnessed a large expansion in the capabilities of observational astronomers to identify new or variable objects in the sky at a variety of wavelengths.  These rapid advances have been made possible largely by the coming of age of wide-field synoptic astronomy, in which a large area on the sky is repeatedly imaged and sophisticated software algorithms search the resulting data stream for new objects or other changes of astronomical interest.

Even in this era, more traditional modes of discovery remain relevant.   Large classical observatories with smaller fields of view often have substantially greater sensitivity and resolution, providing greater depth and less confusion.  While only a small fraction of the observing time on such facilities is spent on projects expressly designed to find new transients or high-amplitude variables, repeat imaging of certain fields (by design or by chance) offers the possibility of finding new and unexpected sources at these locations.  While lacking in the cadence control or blind target selection often employed by dedicated transient surveys, pointed observations can target particularly interesting, exotic, or nearby environments.   These locations may host unusual types of objects that are too rare, in too difficult an environment, or too low-luminosity to be easily identifed in a large-scale synoptic survey.

In this paper we report the serendipitous discovery of a new radio source very close to, but not coincident with, the nucleus of the Cygnus A host galaxy.  Our Karl G. Jansky Very Large Array (VLA) observations leading to the discovery of the transient are described in \S \ref{sec:observations}.  Additional follow-up observations with the VLA and Very Long Baseline Array (VLBA), and archival observations from these and other facilities are also presented.  In \S \ref{sec:discussion} we consider possible physical interpretations for the object, including a rare type of supernova or a fast-accreting secondary supermassive black hole inside Cygnus A.  Finally in \S \ref{sec:implications} we consider the implications of our study for the nature of the Cygnus A system and other luminous radio galaxies, and for radio transient surveys generally.

\section{Observations}
\label{sec:observations}

\subsection{Discovery}
\label{sec:discovery}

Cygnus A is the best-studied powerful radio galaxy by far \citep{Carilli+1996}.  It is the archetype for the \cite{Fanaroff+1974} class II radio galaxies, in which two powerful oppositely-directed jets of relativistic matter are observed to emanate from a central point source at the galaxy nucleus and terminate at bright hot spots in extensive edge-brightened radio lobes in the halo.

Cygnus A was observed by the VLA during the mid-1980s, revealing its inner jet and luminous arcsecond-scale hotspots at the jet termination points \citep{Perley+1984,Carilli+1991}.  The inner core and jets of Cygnus A have been also been studied extensively on milliarcsecond scales using very long baseline interferometry (VLBI) techniques from 1.4 GHz to 90 GHz \citep{Carilli+1994,Krichbaum+1996,Krichbaum+1998,Boccardi+2016b,Boccardi+2016a}.

In spite of or perhaps because of the success of the early VLA observations, no additional sensitive measurements of Cygnus A at $>$10 pc scales were acquired until quite recently.  Motivated by the major improvements to the VLA's bandwidth and sensitivity \citep{Perley+2011}, a new wideband radio-frequency imaging campaign of the system was initiated in 2015, accompanied by a deep \emph{Chandra} X-ray observation.  The radio campaign used all of the VLA's receivers between 2--18 GHz and all four array configurations.

\begin{figure}[t!]
\centerline{
\includegraphics[width=8.5cm,angle=0]{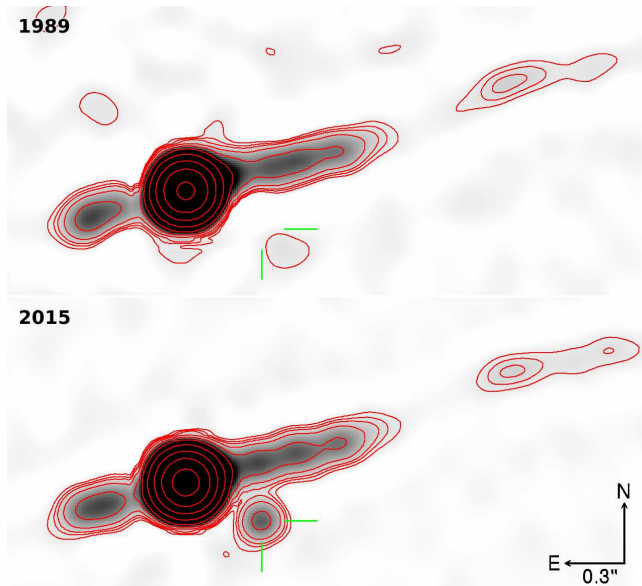}} 
\caption{Discovery images of the off-nuclear radio transient Cygnus A-2.  Both images show VLA observations at 8.5 GHz, using spacings of $>$150k$\lambda$ only.  The scale and contours are the same for both panels (the first five contours are 0.5, 0.85, 1.2, 2, and 4 mJy/beam).  A point source is detected approximately 0.42$\arcsec$ west-southwest of the nucleus in the new observations (location designated by the green crosshairs).  No source at this location was present in 1989.}
\label{fig:discovery}
\end{figure}

Most of the reduced images show similar structure as the original VLA images from the 1980s, with greater sensitivity and wider frequency coverage.  However, in analyzing the higher-frequency extended-configuration observations we noticed a new feature that was not evident in any previously published imaging of the system:  a strong point source (4 mJy at 8 GHz) at an offset of 0.42$\arcsec$ west-southwest of the nucleus (Figure \ref{fig:discovery}).  This is not along the jet axis, but is embedded in the complex and gas-rich inner region of the host galaxy seen in prior optical imaging \citepeg{vandenBergh+1976}.  The source is visible at the same location at multiple frequencies and the detection is highly secure ($>$12$\sigma$ detection at all frequencies), leaving no doubt that it is a real object.  We designate this source Cygnus A-2 (or A-2 for short).

\subsection{VLA Observations}
\label{sec:vla}

To confirm that A-2 represents a new source (rather than a non-variable object that was below the detection limit of the earlier, less-sensitive observations), we searched the NRAO archives for observations taken in configurations and frequencies suitable in principle to resolve and detect a source at this location.  We found two suitable sets of archival observations.  A low-frequency observation of the nucleus was taken on 1989 Jan 06 in X-band (program ID AC244), using four spectral windows (centered at 7815, 8165, 8515, 8885 MHz) with 6.25 MHz of bandwidth each (the narrow bandwidth was necessary to reduce chromatic aberration at the 1-arcminute offset of the source's two hotspots).  In addition, high-frequency observations were obtained  on 1996 Nov 11--12 (in Q-band) and 1997 Mar 29 (in K-band and Q-band), both part of program ID AP334.  These observations were both taken with a subarray using 13 antennas, and with 50 MHz of bandwidth centered at 22.46 GHz (K-band) and 43.34 GHz (Q-band).

We also applied for and received additional VLA observations under director's discretionary time (program IDs 16B-381 and 16B-396) in order to determine if A-2 was still present one year after the discovery observation and, if so, to better constrain its spectrum and rate of evolution.  All frequency bands capable of separately resolving the target from the Cygnus A nucleus in the available configuration were used in these programs: K through Q bands (18--50 GHz) on 2016 Aug 14 in B-configuration, and X through Q bands (8--50 GHz) on 2016 Oct 21 in A-configuration.  All observations, as well as the 2015 discovery observations, were taken using the WIDAR correlator in continuum mode, using 3-bit sampling and the maximum bandwidth available for each receiver.

All data were calibrated in AIPS using well-established techniques.    The flux scale was set by observations of J1331+3030 (3C286), using the flux density scale of \cite{PerleyButler2013}.  Referenced pointing, utilizing observations of the nearby unresolved source J2007+4029, was used to stabilize the antenna pointing on Cygnus A.  We used the Cygnus A nucleus, rather than J2007+4029, to establish the phase calibration: standard phase calibration using this source does not fully remove the differential atmospheric phases present between that source and Cygnus A.  The nucleus can be treated as pointlike at VLA resolutions, as the jet and lobe structures and the hotspots are largely resolved out at spacings beyond 0.5 M$\lambda$.  We employ only these long spacings to establish the phase calibration where possible.  For the high-frequency A-configuration data, the hotspots are completely resolved out (and furthermore lie near the first null of the antenna pattern), so all interferometer spacings were employed.  For the B-configuration data, and for the A-configuration data at the longest wavelengths (C, X, and Ku bands), residual emission from the hotspots was managed using flanking fields in the imaging/deconvolution process.  Following this self-calibration step, the data were decimated into 1\,GHz-wide (for C and X bands) and 2\,GHz-wide (for the other bands) continuum blocks for the imaging stage.

Spectral fluxes for both the nucleus and A-2 were determined using the AIPS task \texttt{JMFIT}.  The individual spectral windows were averaged together into 1 GHz or 2 GHz wide bins; these values are specified in Table \ref{tab:fluxes}.   All measurements were consistent with a point source, with no indication of spatial extension at the VLA's resolution ($\sim$40 mas for the longest spacings at the high frequencies).  Note that the values in Table \ref{tab:fluxes} do not include systematic errors associated with uncertainty in the flux density scale, which are estimated to be a few percent at each frequency \citep{PerleyButler2013}.  A-2 is not well-resolved from the nucleus in the lowest-frequency 2016 B-configuration spectral window or in any of the 2015 A-configuration observations below 7 GHz, so we do not quote fluxes at these frequencies.

The archival observations in 1989 and 1996--1997 unambiguously show that A-2 was not present then to limits substantially below its discovery value.  Using the RMS of the synthesized map at locations close to the position of A-2, we place a limiting flux of $<$0.69 mJy at 8.34 GHz (1989), $<$1.32 mJy at 22 GHz (1996), and $<$0.78 mJy at 45 GHz (1997).  This indicates that the flux rose by at least a factor of six between 1989 and 2015 (and at least four between 1997 and 2015).

No obvious variability is seen during the past year.   Comparing repeat measurements at the same frequency (i.e., comparing our October 2016 observations to the low-frequency points from 2015 and high-frequency points from August 2016), no measurement changes by more than 5 percent and only one measurement changes by greater than 3$\sigma$.  (However, the noise structure close to the nucleus is complex, and a 3$\sigma$ change of this individual point does not securely indicate real variability).  There is some indication that the flux may be dropping slightly (at least at the lower frequencies where a longer time baseline is available), but any such behavior is marginally significant at best and at the level of no more than a few percent per year.   A longer temporal baseline will be required to determine more clearly if the object is indeed changing with time.

\begin{figure}[t!]
\centerline{
\includegraphics[width=9cm,angle=0]{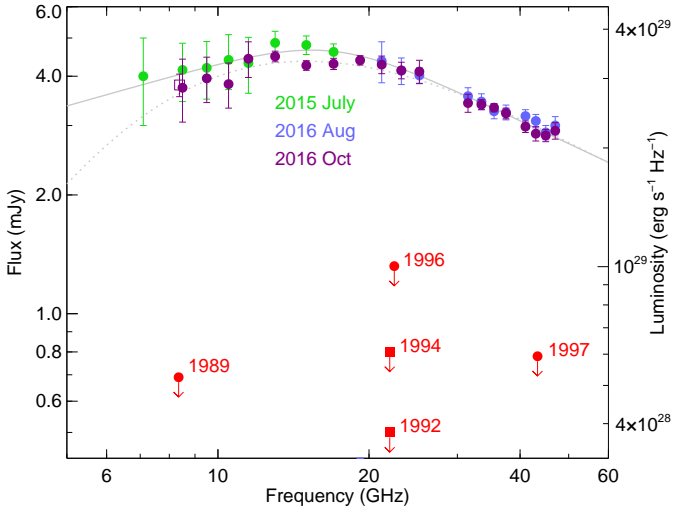}} 
\caption{Spectral energy distribution of the off-nuclear transient (A-2) as measured from three recent VLA epochs (colored circles) and one VLBA epoch (open square).  Upper limits from archival observations using the VLA (red circles) and using high-frequency VLBI (red squares; from Krichbaum et al., private communication) are also presented.  Error bars are 2$\sigma$.  We detect no obvious variability at any frequency over a 1-year baseline, although the source rose by a factor of $\gtrsim$5 between the mid-1990s and 2015.  Two different models are shown, one with a self-absorbed turnover at low frequencies $\alpha=5/2$ (dashed line) and a fully optically-thin model with a gentler turnover (solid line).}
\label{fig:sed}
\end{figure}

A spectral energy distribution (SED) formed by our measurements is presented in Figure \ref{fig:sed}.  The nature of the multiwavelength SED is most unambiguously constrained by our A-configuration observation from October 2016 in which all frequencies are available simultaneously: this shows a largely flat (in $F_\nu$) overall spectral shape, with some curvature at the low-frequency end.  The high-frequency data points ($>20$ GHz) are well-fit by a power-law with a spectral index of $\alpha_{\rm hi}$=$-$0.6$\pm$0.1 (using the convention $F_\nu \propto \nu^{\alpha}$).    Below 15 GHz the SED flattens, and probably drops at the lowest frequencies.  The low-energy spectral index is poorly constrained and depends largely on what is assumed about the softness of the spectral turnover.  The best fits are obtained with a sharp turnover and a rather flat low-frequency index ($\alpha_{\rm lo} \approx +0.4$) but steep low-frequency indices (including the self-absorbed value of $\alpha=+2.5$; see discussion in \S \ref{sec:sed}) are permitted if the turnover is soft.

\begin{deluxetable*}{c|ccc|ccc|ccc}  
\tabletypesize{\small}
\tablecaption{Jansky VLA Fluxes of the Cygnus A Nucleus and Off-Nuclear Transient A-2}
\tablecolumns{10}
\tablehead{
\colhead{} &
\colhead{} &
\colhead{2015-07} &
\colhead{} &
\colhead{} &
\colhead{2016-08} &
\colhead{} &
\colhead{} &
\colhead{2016-10} &
\colhead{}
\\
\colhead{$\nu$ \tablenotemark{a}}  &
\colhead{$F_{\nu,{\rm nuc}}$ \tablenotemark{b} } &
\colhead{$F_{\nu,{\rm off}}$ \tablenotemark{c} } &
\colhead{$\sigma$ \tablenotemark{d}} &
\colhead{$F_{\nu,{\rm nuc}}$ \tablenotemark{b} } &
\colhead{$F_{\nu,{\rm off}}$ \tablenotemark{c} } &
\colhead{$\sigma$ \tablenotemark{d}} &
\colhead{$F_{\nu,{\rm nuc}}$ \tablenotemark{b} } &
\colhead{$F_{\nu,{\rm off}}$ \tablenotemark{c} } &
\colhead{$\sigma$ \tablenotemark{d}}
\\
\colhead{(GHz)} &
\colhead{(mJy)} &
\colhead{(mJy)} &
\colhead{(mJy)} &
\colhead{(mJy)} &
\colhead{(mJy)} &
\colhead{(mJy)} &
\colhead{(mJy)} &
\colhead{(mJy)} &
\colhead{(mJy)}
}
\startdata
 7.1  & 1393 & 4.0   &  0.50&    -  &  -    &    -   &    -  &   -    &    - \\
 8.5  & 1368 & 4.15  &  0.35&    -  &  -    &    -   &  1253 & 3.74   &  0.34 \\
 9.5  & 1416 & 4.20  &  0.35&    -  &  -    &    -   &  1283 & 3.95   &  0.26 \\
10.5  & 1483 & 4.40  &  0.35&    -  &  -    &    -   &  1399 & 3.82   &  0.25 \\
11.5  & 1507 & 4.32  &  0.35&    -  &  -    &    -   &  1423 & 4.43   &  0.23 \\
13.0  & 1440 & 4.86  &  0.17&    -  &  -    &    -   &  1414 & 4.49   &  0.07 \\
15.0  & 1435 & 4.80  &  0.13&    -  &  -    &    -   &  1428 & 4.26   &  0.06 \\
17.0  & 1427 & 4.61  &  0.11&    -  &  -    &    -   &  1450 & 4.30   &  0.07 \\
19.2  &   -  &  -    &   -  &    -  &  -    &    -   &  1498 & 4.39   &  0.06 \\
21.2  &   -  &  -    &   -  &  1527 & 4.37  &  0.26  &  1498 & 4.28   &  0.11 \\
23.2  &   -  &  -    &   -  &  1505 & 4.13  &  0.16  &  1475 & 4.14   &  0.10 \\
25.2  &   -  &  -    &   -  &  1438 & 4.02  &  0.09  &  1473 & 4.11   &  0.14 \\
31.5  &   -  &  -    &   -  &  1320 & 3.56  &  0.09  &  1317 & 3.42   &  0.09 \\
33.5  &   -  &  -    &   -  &  1280 & 3.45  &  0.08  &  1276 & 3.39   &  0.05 \\
35.5  &   -  &  -    &   -  &  1220 & 3.26  &  0.07  &  1248 & 3.33   &  0.04 \\
37.5  &   -  &  -    &   -  &  1199 & 3.24  &  0.07  &  1201 & 3.22   &  0.05 \\
41.0  &   -  &  -    &   -  &  1198 & 3.17  &  0.06  &  1203 & 2.98   &  0.05 \\
43.0  &   -  &  -    &   -  &  1176 & 3.08  &  0.06  &  1156 & 2.86   &  0.06 \\
45.0  &   -  &  -    &   -  &  1158 & 2.88  &  0.06  &  1133 & 2.83   &  0.05 \\
47.0  &   -  &  -    &   -  &  1141 & 3.00  &  0.08  &  1114 & 2.91   &  0.07 \\
\enddata
\label{tab:fluxes}
\tablenotetext{a}{\ Central frequency of the observation.}
\tablenotetext{b}{\ Flux density of the Cygnus A primary nucleus.}
\tablenotetext{c}{\ Flux density of A-2, the variable off-nuclear point source.}
\tablenotetext{d}{\ Uncertainty of the flux density of A-2, excluding calibration uncertainties.}
\end{deluxetable*}

\subsection{VLBA Observations}
\label{sec:vlba}

We also acquired director's discretionary observations with the Very Long Baseline Array (program ID BP213).  Observations were taken in the S- and X-bands simultaneously in dual polarisation, dual band with dichroic, a 2 Gbps total recorded bitrate, using the DDC (digital down-converter) signal processing mode. The recording was split equally between S-band (using 2$\times$64 MHz of bandwitdth centered at 2230 and 2345 MHz) and X-band (using 2$\times$64 MHz of bandwidth centered at 8350 and 8420 MHz).  We used filler time at low priority, with 9 or 10 antennas, in each of four observations.  The observations were taken on 2016 Nov 3, Nov 11, Nov 14, and Nov 20.  The total on-source time was 3.1 hours.  

We observed continuously on Cygnus A, and used the strong nucleus to calibrate delay and phase for the observation. The transient (A-2) is well within the primary beam of the antennas and experiences the same phase and delay variations from the atmosphere, which are hence almost completely removed by calibration.    Data reduction followed the standard path for VLBA data in AIPS, i.e., amplitude calibration by applying the pre-measured system temperature and antenna efficiency factors provided by NRAO operations, followed by delay, rate, and phase calibration.

Only the X-band data (Stokes I) were usable.  (The S-band observations were unsuccessful due to a combination of factors, including radio-frequency interference, interstellar scattering, and instrumental challenges associated with the luminous lobes within the primary beam; no image could be produced.)  Even for the X-band data, the flux on the longest baselines is weak due to a combination of intrinsic structure and/or scattering, so we tapered the final images to 100 M$\lambda$ (from 200 M$\lambda$) to down-weight low SNR data.

\begin{figure}[t!]
\centerline{
\includegraphics[width=7.5cm,angle=0]{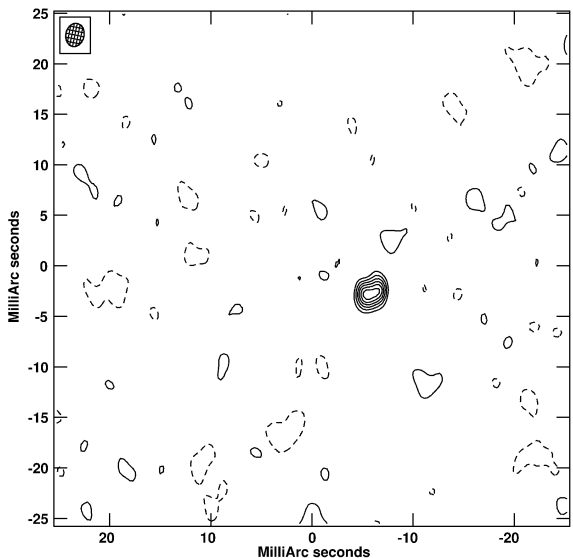}} 
\caption{VLBA synthesized image of the A-2 field.  A pointlike source with flux consistent with the VLA measurement is detected at a location consistent with the VLA (and NIR) locations, indicating that the source is quite compact ($<4$ pc). Brightness contours are nonzero integer multiples of 0.47 mJy/beam.}
\label{fig:vlba}
\end{figure}

Our final reduced image of the system, combining all four epochs, has an RMS of 0.25 mJy and a final beam size of 2.3$\times$1.8 mas (FWHM) and is shown in Figure~\ref{fig:vlba}.  The new source A-2 is clearly detected, with an integrated flux of 3.8$\pm$0.25 mJy and peak brightness of 3.1$\pm$0.25 mJy/beam.  Some possible east-west extension is apparent in the image, and formally the FWHM of the source along the major axis is broader than the synthesized beam at 3$\sigma$ significance, hinting that A-2 may be marginally resolved on a scale of $\sim$1--2 mas.  However, this extension is marginal and could result from phase-transfer artifacts associated with the use of the primary nucleus for phase calibration.  More conservatively, we place an upper limit on the spatial scale of the source by measuring the full-width of the projected profile in the synthesized image.  We derive a limit on the diameter of $<$4 mas, equivalent to 4.5 pc at the distance of this system ($z=0.0561$; we assume the standard cosmology of \citealt{Bennett+2014}).

The precise location of A-2 is $\alpha$=19:59:28.32345, $\delta$=+40:44:01.9133 (J2000, $\pm$1\,mas), referenced to the previously known astrometric position of the nucleus (established to be $\alpha_{\rm nuc}$=19:59:28.35648, $\delta_{\rm nuc}$=+40:44:02.0963, \citealt{Gordon+2016}) in the ICRF2 frame \citep{Fey+2015}.  The projected offset from the nucleus is 418 mas, or 458 pc.

\subsection{Multifrequency Archival Observations}
\label{sec:oir}

The Cygnus A nucleus is very luminous at all electromagnetic wavelengths from radio to X-rays ($\sim10^{44}$ erg\,s$^{-1}$; \citealt{Carilli+1996}).  The 0.42$^{\prime\prime}$ projected offset places A-2 well within the PSF of the nucleus at the typical resolution of nearly all existing observatories, meaning that only a few facilities (those capable of resolutions less than approximately 0.3$\arcsec$) are capable of detecting it.  At the present time, this is limited to radio observatories, HST, and ground-based near-infrared adaptive optics instruments.

Cygnus A was the subject of several VLBI monitoring campaigns in the early 1990s.  These data are not publicly archived, but we contacted the authors of the most recent large-scale VLBI study \citep{Krichbaum+1998} who re-investigated their 22 GHz observations for evidence of any emission at the location of A-2.  There is no detection of any source to upper limits of $<$0.5 mJy on 1992 June 10 and $<$0.8 mJy on 1994 March 4.

Cygnus A has also been the target of several HST campaigns, beginning with the study of \cite{Jackson+1994} shortly following the first servicing mission, with additional follow-up studies by \cite{Jackson+1998} and \cite{Tadhunter+1999,Tadhunter+2000,Tadhunter+2003}.  The observations used in these studies were all taken between 1994 and 2001.  These images reveal in detail the inner region surrounding the nucleus, which is dominated by a biconic structure that surrounds the two jets.  No distinct source is evident at the position of A-2 in the blue and UV observations, which is unsurprising given the very large extinction towards the central part of the Cygnus A host galaxy (as well as through the plane of our own Galaxy).  A point source begins to emerge at the location of A-2 redward of approximately 500 nm and is quite distinct in the 814 nm image.  It is evident in the near-IR NICMOS images also, but is difficult to cleanly resolve from its complex environment due to the lower resolution of that camera.  We searched the HST archive to determine if the source has been observed since 2001 and did not find any constraining imaging.

\begin{figure*}
\epsscale{1.1}
\plottwo{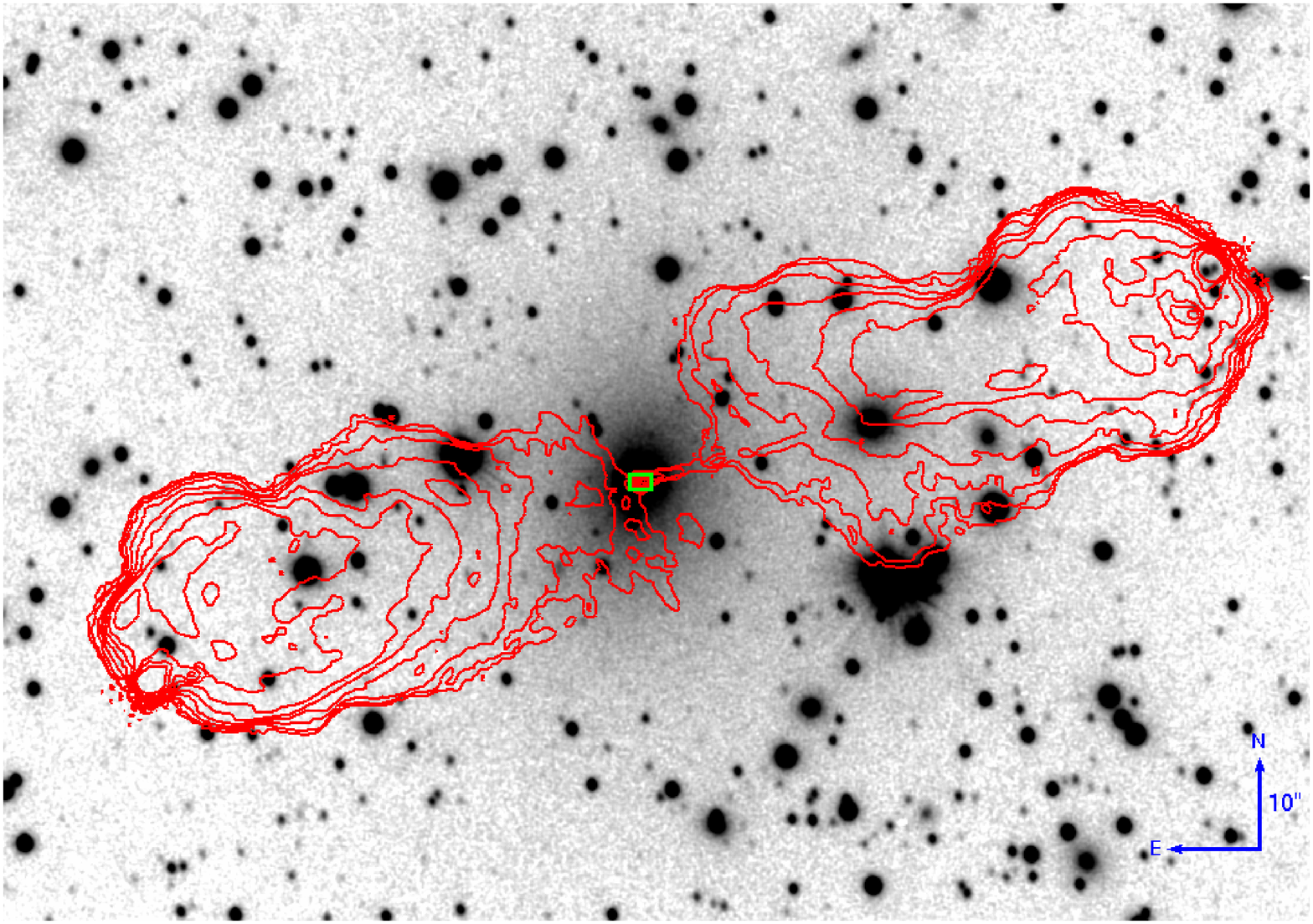}{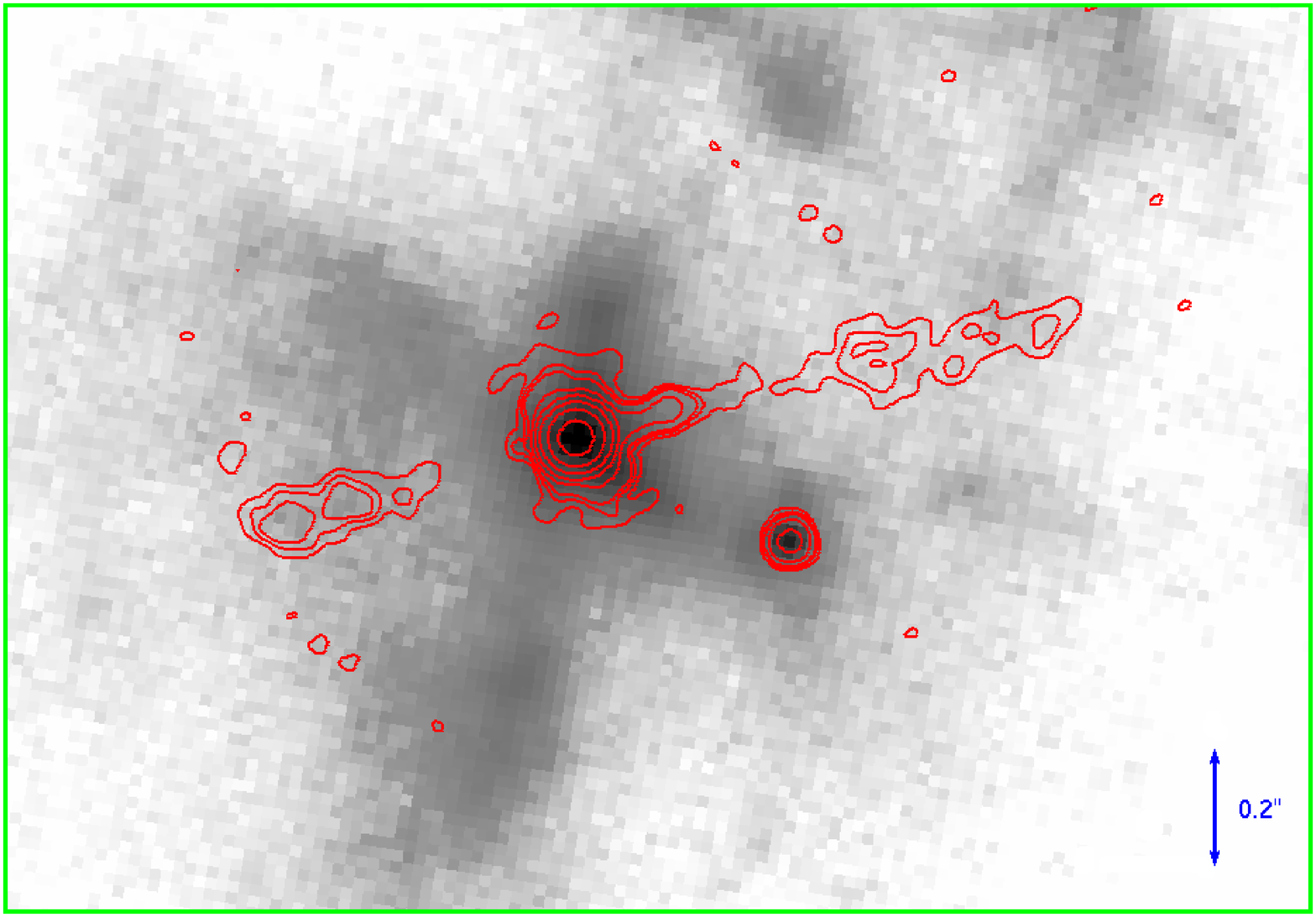}
\caption{Multiwavelength imaging of the new source in Cygnus A.  At left we show a wide-field image of the Cygnus A region; the grayscale background image is optical imaging from Gemini while the contours are VLA observations at 2 GHz from November 2015, demonstrating the iconic jet and lobe structure.   At right a zoom-in on the nuclear region is shown, with the grayscale from the Keck AO imaging of \cite{Canalizo+2003} and radio contours measured from a 35 GHz VLA image taken in October 2016.  A distinct, luminous point source is detected 0.42$\arcsec$ from the luminous nucleus in the radio band and in the NIR imaging.}
\label{fig:image}
\end{figure*}

Cygnus A was also observed using Keck adaptive optics (NIRC2) in May 2002 by \cite{Canalizo+2003}.  Observations were acquired in all three near-infrared bands ($J$, $H$, and $K'$); these unambiguously show a strong point source at the location of A-2 (Figure \ref{fig:image}).  This source is discussed extensively in that work; the authors favor an interpretation in which it represents the dense, stripped stellar core of a companion galaxy merging into the Cygnus A host.   We were provided the original $K'$-band AO image by C. Max and aligned it with our highest-frequency A-configuration image from the VLA.  Registering the radio and IR images using the nucleus, the positional coincidence of the point source with A-2 in both images is precise to within 0.01$\arcsec$ or 10 pc, indicating a secure physical association between the two objects.  

All of the HST and AO imaging of this field was single-epoch, and to our knowledge no re-observations of this source at a common waveband have been conducted to date.  The rate of change of flux of the source is therefore not strongly constrained.  Assuming no change between the different HST epochs this optical counterpart must be very red, as would be expected for a source embedded within the Cygnus A inner environment.  It is however significantly less red than the nucleus itself, so the extinction column towards A-2 is likely somewhat lower (the nucleus is invisible at all optical wavelengths but the optical/IR counterpart of A-2 can be detected down to $\sim$500 nm).

\section{Interpretation}
\label{sec:discussion}

\subsection{Association with the Cygnus A Host Galaxy}

We consider first the possiblity that the variable radio source simply represents a coincidental object unassociated with Cygnus A, such as an active M-dwarf in the disk of our own Galaxy.   The probability of this is extremely low.  Even at low Galactic latitude ($b=5.75\arcdeg$), the probability that a detectable Galactic field star with $K < 20$ mag would appear within an arcsecond of the nucleus of this galaxy is quite low, $<10^{-4}$.  (The possibility of a non-Galactic point source such as a quasar being coincidentally aligned this way is orders of magnitude lower.)  The foreground/background hypothesis was rejected on these grounds alone by \cite{Canalizo+2003}, who furthermore noted that the colors of the object are not consistent with typical Galactic stars.   The radio detection further strengthens the statistical case for an extragalactic origin, since it is very unlikely that a random foreground star would also be radio-loud.  We therefore are quite confident that the source originates in Cygnus A.

\subsection{Luminosity Constraints}
\label{sec:lum}

Placing the source at the redshift of Cygnus A ($D_L=251$ Mpc) allows us to calculate its luminosity: $L_\nu \approx 3 \times 10^{29}$ erg s$^{-1}$ Hz$^{-1}$ or $\nu F_\nu \approx 6\times10^{39}$ erg s$^{-1}$.  This alone represents an extremely powerful constraint, since it is orders of magnitudes more luminous than almost any known variable radio object.  It rules out, in particular, any known nondestructive stellar event (such as a flare, a nova, or any known class of energetic burst from a pre-existing compact stellar object such as a neutron star or stellar-mass black-hole binary): see, for example, the luminosity-timescale diagram of Figure 3 in \cite{Pietka+2015}.

Viable explanations consistent with the luminosity of A-2 and its appearance during a $\lesssim$10-year timescale can be sorted into two general categories: an exceptionally luminous class of supernova, or a rapidly accreting supermassive black hole.  We will compare these two models in more detail in \S\ref{sec:sn} and \S\ref{sec:smbh} after examining a few additional physical considerations.

\subsection{Size Constraints}
\label{sec:size}

The VLBA observation directly constraints the maximum size of the object to be less than approximately 4 pc.  However, we can also place a \emph{lower} limit on the size on account of the lack of strong variability: given the low Galactic latitude of Cygnus A, for a compact source we would expect to observe large-amplitude interstellar scintillations due to refraction by ionized gas within the plane of the Milky Way.  

The transition frequency for the strong interstellar scintillation regime is quite high in this direction (approximately 30 GHz according to the maps of \citealt{Walker+2001}).  At frequencies close to this value, large (of order unity) modulations of the observed flux are to be expected for sources smaller in angular size than the Fresnel scale in this direction, which is $\sim 1$\,$\mu$as or $\sim10^{-3}$\,pc in projection \citep{Walker+1998}; the variation timescale is approximately 1 day.  Although we only have three epochs at this time, our observations are all close to the critical frequency and it is very unlikely that even three repeated observations on timescales far greater than the characteristic fluctuation time would provide consistent measurements within a few percent.  This suggests that the source is larger (most likely significantly larger) than $10^{-3}$ pc (200 AU).

\subsection{SED Constraints}
\label{sec:sed}

The SED of A-2 is nonthermal throughout the observed bands, as would be expected given the very high brightness temperature ($T_B > 1.7\times10^7$\,K at 8 GHz, given the flux measurement and angular size limit provided by the VLBA observations).  This implies a population of highly energetic particles, most likely shocked or otherwise relativistically accelerated electrons radiating via synchrotron emission in a local magnetic field, the same process that is responsible for the radio emission from nearly all energetic transient phenomena (including both SNe and AGNs).  The observations are consistent with this; the expected high-frequency spectral index for a shocked synchrotron population is $\alpha = -(p-1)/2$, which for the observed $\alpha=-0.6$ (\S \ref{sec:vla}) implies $p=2.2$, a standard value for this term.

The observed spectral energy distribution appears to roll over below $\nu_t \approx 15$\,GHz.  The lack of short-term variability establishes that this is not likely to be due to scintillation (\S \ref{sec:size}).  A spectral turnover could in principle also originate from free-free absorption by ions along the line of sight through the host galaxy.  However, the required emission measure to produce a turnover via free-free emission is enormous  ($EM \approx 1.5 \times \nu_{t,{\rm MHz}}^{2.1}$ pc\,cm$^{-6}$ $\approx$ $8.8 \times 10^8$ pc cm$^{-6}$ for $\nu_t$=10 GHz; \citealt{Condon+2016}), a factor of $\sim10^7$ times higher than what has been seen in narrow-band H$\alpha$ observations of the galaxy \citep{Carilli+1989}.  And while A-2 is quite optically obscured, the detection of a clear optical/IR counterpart at this location (\S \ref{sec:oir}) demonstrates that the extinction along this sightline cannot be too extreme (assuming an intrinsically flat SED in $f_\nu$, the colors provided by \citealt{Canalizo+2003} imply an extinction factor of no more than $\sim$50 at the wavelength of H$\alpha$).   Alternatively, the absorbing matter could have been recently ionized and thus not evident since the time of the most recent H$\alpha$ observations---as would be the case in a Type IIn supernova model, in which the progenitor star releases a dense wind and then ionizes it upon explosion \citep{Chevalier+1982}.  This would require a luminous UV/optical transient to have accompanied the appearance of A-2 and predict strong and ongoing H$\alpha$ emission from the transient; possibilities that could be checked via archival and/or future observations, respectively.

Alternatively, synchrotron self-absorption is commonly seen in extragalactic sources and is generically expected at low frequencies anytime that the accelerated electrons are confined to a relatively limited volume.  The observed self-absorption frequency $\nu_{\rm SSA}$ is related to the source size $R$ according to the expression: $R \approx 9$\,$\times$\,$10^{15}$\,${\rm cm} \times (\frac{F_{\rm peak}}{\rm Jy})^{\frac{9}{19}}(\frac{D_L}{\rm Mpc})(\frac{\nu_{\rm SSA}}{\rm 5 GHz})^{-1}$ (\citealt{Chevalier+1998}; terms of order unity are omitted).   For our observed parameters this corresponds to a size of $R \approx 0.1$\,pc, fully consistent with the size constraints we have derived above (\S \ref{sec:size}).  (If the break is not due to SSA, then this estimate becomes an inequality: $R>$0.1\,pc).   

A spectral break could also originate from the acceleration process itself: for example, if all of the accelerated electrons exceeded a certain minimum (large) Lorentz factor, as is the case for extremely energetic shocks such as those in gamma-ray bursts \citepeg{Sari+1998}.   Lower-frequency observations will be needed to determine the low-frequency spectral index (expected to be $\alpha=2.0$ for free-free absorption, $\alpha=2.5$ for synchrotron self-absorption, or $\alpha = 0.33$ for electron injection) and confirm the nature of the break.

\subsection{Supernova Models}
\label{sec:sn}

It is very rare for a supernova to reach the radio luminosity we observe for A-2 \citep{PerezTorres+2015}.  Type Ia SNe are never detected in the radio band, and common core-collapse supernova classes reach only $10^{26}$\,erg\,s$^{-1}$\,Hz$^{-1}$ (IIp supernovae) to $10^{28}$\,erg\,s$^{-1}$\,Hz$^{-1}$ (most Ib/c supernovae).  However, some exotic classes of supernova can reach the luminosity scales in question.  

\emph{Relativistic} supernovae---a subclass of Ib/c events---entrain a significant amount of energy in jets traveling at relativistic velocities.  The rarest and most luminous of these, long-duration gamma-ray bursts (GRBs), have extremely powerful on-axis jets whose interaction with their environments creates powerful afterglows with radio luminosities exceeding $10^{31}$\,erg\,s$^{-1}$\,Hz$^{-1}$ at peak and lasting for years \citep{Chandra+2012}.  The more common mildly-relativistic versions may or may not possess strong narrow jets, but do accelerate a significant amount of matter to near $c$; examples include SN 2009bb or PTF 11qcj, which exhibited maximum radio luminosities around 10$^{29}$\,erg\,s$^{-1}$\,Hz$^{-1}$ \citep{Soderberg+2010,Corsi+2016}.

\emph{Strongly interacting} (type IIn) supernovae are not relativistic but owe their luminosity to the collision of the expanding supernova envelope with massive, dense circumstellar matter ejected by the star prior to explosion.  Only a few of these have been well-studied in the radio, but their luminosities are comparable to mildly-relativistic type Ic supernovae.

It would not be surprising to identify a supernova within the inner environs of Cygnus A.  The star-formation rate is very large, perhaps as high as 80 M$_\odot$ yr$^{-1}$ \citep{Privon+2009,Hoffer+2012}, equivalent to a core-collapse supernova rate of approximately 0.7 yr$^{-1}$ \citep{Horiuchi+2011}, so during any given observation it is likely that there are several young supernovae within a few years of explosion.  However it is quite \emph{un}likely that we would find a young \emph{radio-luminous} supernova by sheer chance if star-formation in the Cygnus A host is similar to that in other galaxies.  Cygnus A-2 would have to be \emph{the} most radio-luminous non-GRB supernova ever recorded.  Broad-lined SNe-Ib/c and SNe-IIn combined represent less than 5\% of the core-collapse supernova rate in typical galaxies (\citealt{Arcavi+2010,Graur+2016}; true GRBs are orders of magnitude rarer still), and radio surveys of these classes indicate that less than 10\% of either could be as luminous as our source \citep{Soderberg+2006,Corsi+2016}.  The chance that a blind observation of Cygnus A would identify an extremely radio-luminous SN within 10 years after explosion is therefore $\lesssim 5 \times 10^{-3}$.

This is far from impossible, and furthermore it is quite possible that gas-rich, high-density environments such as the Cygnus A nucleus may produce a different distribution of supernovae than more mundane star-forming environments.  The most luminous known (probable) interacting radio supernova was also located in a dusty nuclear region around an AGN (Markarian 294A), for instance \citep{Yin+1994}, and two examples of optically superluminous type IIn supernovae, SN\,2006gy and PTF\,10tpz, have also been found in the nuclear environments of massive galaxies, so it is possible that these environments are particularly friendly to very massive stars and energetic supernovae \citep{Perley+2016}.  

However, despite being quite nearby (77 Mpc) SN\,2006gy was undetected by the VLA \citep{Ofek+2007}, and no radio detection of any other (optically) superluminous SN IIn has been reported to date to our knowledge.  Also, larger-scale VLBA radio surveys of ULIRGs with dense, dusty star-forming nuclear environments similar to what is present in Cygnus A have not found clear evidence of an excess number of very luminous supernovae compared to what is expected given their high star-formation rates \citep{Lonsdale+2006}.  Furthermore, optical surveys suggest that (if anything) the relative fractions of IIn and broad-lined Ib/c supernovae actually \emph{decrease} in the most massive and metal-rich galaxies \citep{Arcavi+2010,Modjaz+2011,Graur+2016}, as do GRBs \citepeg{Perley+2016a,Japelj+2016,Graham+2017}.

A supernova model also is unnatural given our observations of the system to date.  GRBs are the only stellar transients actually observed to have exceeded the observed luminosity of our source---but a GRB at this location would have to be many years old to not be varying at any frequency over a timescale of one year, in which case its luminosity would be surprising even for a gamma-ray burst.  An old ($>$5 years), highly relativistic event would likewise be expected to have become resolved in VLBA imaging by this point, given the anticipated superluminal lateral expansion of the jet \citep{Taylor+2004}.

Another serious problem with supernova models is the precise coincidence with an optical/IR point-source, which is unexpected and difficult to explain.\footnote{While the optical luminosity of the counterpart of A-2 ($m_K\approx-18$) is similar to that of SNe, it cannot be a supernova itself, as the point source is visible in both HST images from the mid-1990s and the Keck observations from 2002.}  This could represent a young and very dense star cluster (compact and luminous star clusters are known in the central Milky Way and in other luminous galaxies), with the Cygnus A transient representing one of the first supernovae from an extremely massive star inside of it.  However, the narrow-band optical and spectroscopic NIR observations of this source show no evidence that it produces strong line emission, as would be expected from a young super star-cluster \citep{Jackson+1998,Canalizo+2003}.   Also, given that the probability of catching an ultra-rare supernova by chance in this galaxy is very low to begin with, the possibility that said SN would \emph{also} happen to align with a particular single cluster and not originate in one of the many other abundant star-forming regions within Cygnus A makes this statistical difficulty even more problematic.

Further radio monitoring will be able to examine the supernova hypothesis more robustly.  A multi-year radio light curve should unambiguously show fading for any GRB-like model (and likely for a IIn model as well), higher-frequency VLBA observations will place tighter constraints on jetted emission, and further adaptive optics or HST observations with modern instruments may be able to identify optical or NIR line emission from a late-stage nebular supernova.   In the meantime, however, we will focus our attention on the alternative black hole model, for which a high radio luminosity and coincidence with a bright optical point source are natural expectations.

\subsection{Accreting massive black hole models}
\label{sec:smbh}

As Cygnus A-2 is clearly separate from the nucleus that powers the Cygnus A jets it cannot be related to the primary supermassive black hole in this galaxy.  However, there is no reason that a galaxy such as this one cannot harbor more than one large black hole.   Indeed, although most giant elliptical galaxies at low redshift are starved of gas and largely devoid of star-formation, the extremely dust- and gas-rich nucleus of Cygnus A (and its rapidly-accreting active nucleus) suggests that a recent merger has delivered a large gas supply to the center of the galaxy in the relatively recent past.  It seems quite plausible that the central supermassive black hole of this infalling satellite has not yet merged with the primary black hole.  Accretion of gas onto such a secondary could produce luminous jets of relativistic matter, leading to variable multiwavelength emission on a variety of wavelengths and timescales.  This could naturally produce a radio transient on a timescale of several years as well as the persistent optical/NIR emission seen in pre-existing observations.

\subsubsection{Secondary AGN}

Our available constraints on the SED, luminosity, size, and variability timescale of Cygnus A-2 are all consistent with what have been observed previously from AGNs \citepeg{Ho+2008}.  An AGN model is also consistent with the short-wavelength observations: the optical/IR counterpart of A-2 could represent either the stripped remnant core of the merging galaxy (as originally proposed by \citealt{Canalizo+2003}) or continuum emission from the AGN itself.  The $K$-band near-infrared spectrum of this object (Figure 5 of \citealt{Canalizo+2003}) does indeed appear quite similar to known AGNs (e.g. \citealt{Glikman+2006,RamosAlmeida+2009}), although beacuse it is heavily contaminated by light from gas excited by Cygnus A itself it is not clear if these features are intrinsic to the point-source (\citealt{Canalizo+2003} note that the centroid of the line emission is offset from the centroid of the continuum).  If AGN-dominated, the bolometric luminosity inferred from the NIR observations ($\nu F_\nu \sim ~2\times10^{41}$ erg s$^{-1}$) is much less than Eddington for a central supermassive black hole (e.g., $\sim$0.002 $L_{\rm edd}$ for $M_{\rm BH} = 10^6\,{\rm M}_\odot$ and an assumed efficiency of $\eta=0.1$), similar to low-luminosity AGNs.  Our unresolved VLBA observation rules out any highly relativistic jet emission, but the more typical mildly-relativistic low-luminosity AGNs are consistent with remaining point-like on this scale.

The most distinctive property of A-2 is its rapid appearance: from nondetection to a clear detection over a timescale of less than nine years.  Most AGNs vary in flux at some level but do not turn on (or off) completely.  Of course, a varying source can move up and down across a detection threshold, and it is entirely possible that the source was present (and accreting) in the 1980s and 1990s at a lower level before brightening over the past decade.  Unfortunately the limited sensitivity of 20th century radio facilities makes it difficult to constrain this precisely.  A change by a factor of six is very large for a non-blazar AGN; for example, of the blindly-selected radio sources in Stripe 82 of the Sloan Digital Sky Survey examined by \cite{Hodge+2013}, only 1\% showed variability by more than a factor of a few on a decade timescale.  It seems curious that, should a secondary AGN be present in Cygnus A, that it would happen to be one of these strong variables.  On the other hand, A-2 was discovered in a very different manner than that by which ordinary AGNs are selected and in an atypical environment, so the statistics on short-term variability based on blindly-selected AGNs may not be representative.  Also, sudden changes in AGN fluxes and spectra at a variety of wavelengths have recently been seen in some well-studied nearby AGNs, including NGC\,2617 \citep{Shappee+2014}, Mrk\,590 \citep{Koay+2016}, and NGC\,660 \citep{Argo+2015}, although only in the lattermost of these examples was a sharp increase observed at radio wavelengths.

\subsubsection{Tidal disruption event}

Alternatively, a largely or completely quiescent AGN could suddenly become quite luminous as a result of a large-scale accretion event.  The most extreme example of this would be the disruption and accretion of a star passing near the tidal radius (a tidal disruption event or TDE).  A TDE model is also consistent with the available observational data; some TDEs have luminosities comparable to what we have observed for A-2 and this emission can persist for several years \citepeg{Zauderer+2011}.  However the expected TDE rate per galaxy is very low ($\sim10^{-4}$ yr$^{-1}$): as with the rare classes of supernovae discussed in the previous section, the probability that we would catch such an event by chance is quite low \emph{unless} Cygnus A has an atypically high disruption rate.   There is some reason to expect that this may indeed be the case:  nearly all known TDEs have been localized to galaxies that have undergone recent mergers \citep{Arcavi+2014,French+2016,Lezhnin+2016}, although these systems appear to be in much later stages in which the nuclear black holes have most likely already merged.  The actual TDE rate in ongoing massive galaxy mergers similar to Cygnus A is very difficult to determine observationally, but the detection of a candidate TDE in a nearby ULIRG gives some reason to think that it may indeed be extremely high \citep{Tadhunter+2017}.

\subsubsection{Mass constraints and IMBH/ULX}

Direct observational constraints on the mass of the accreting black hole responsible for A-2 in the SMBH scenario are relatively weak.  The lack of any reported large-scale dynamical disturbances close to this location from IR spectroscopy suggests that any black hole there must be significantly smaller than that of the primary Cygnus A SMBH ($M_{\rm primary} \sim 3\times10^9\,{\rm M}_\odot$; \citealt{Tadhunter+2003})\footnote{This may not apply if A-2 originates from a fast-recoiling black hole associated with a previous merger \citepeg{Blecha+2008}.  As no other recoiling black holes are known, however, this seems unlikely in comparison to simply attributing the A-2 black hole to the central black hole of the galaxy responsible for the present merger event.}.  The pre-flare radio non-detection would imply an upper mass limit of $M_{\rm BH}\lesssim 10^8\,{\rm M}_\odot$ via the AGN $L_{\rm radio}-M_{\rm BH}$ correlation of \citep{Franceschini+1998}---however, more recent work does not confirm this correlation \citep{Nyland+2017}, and it would not apply to the TDE scenario in any case.

The lower limit on the black hole mass is likewise not tightly bounded.  If we assume that the black hole is the central BH of a merging companion galaxy then its mass is likely to be at least $>10^5\,{\rm M}_\odot$ (\citealt{Reines+2015}; see also \citealt{Nyland+2012,Nguyen+2017}), especially if the companion was a massive galaxy responsible for delivering large amounts of gas to the Cygnus A nuclear region and triggering the ongoing starburst.\footnote{The black hole could also be a pre-existing ``stalled'' SMBH delivered by an earlier minor merger \citep{Dosopoulou+2017}.}  In principle, however, our observations permit it to be smaller: Eddington-limit arguments based on the bolometric NIR luminosity limit the mass only to $M_{\rm BH} \gtrsim 2\times10^3\,{\rm M}_\odot$ (and no limit can be placed at all if the optical/IR counterpart's luminosity is stellar in origin.)

An intermediate-mass black hole (IMBH) of this type could be pre-existing, or could have formed during the ongoing starburst in the center of a dense, young star cluster (but c.f.\ the spectroscopic constraints on line emission from young stars in \S \ref{sec:sn}).  No IMBHs have yet been clearly established to exist in the low-redshift universe, although ultra-luminous X-ray sources (ULXs) represent a possible candidate.  ULX's are X-ray flaring events that have been hypothesized to be associated with accretion onto intermediate-mass black holes (IMBHs) in nearby star-forming galaxies \citep{Mezcua+2011}.  However, known ULX flares have radio luminosities orders of magnitude lower than A-2 \citep{Mezcua+2011,Webb+2012,Pietka+2015}, so if A-2 originates from an IMBH flare this event would have to be without observational precedent.  We conclude that is much more likely that the A-2 black hole is supermassive and originates from the center of a merging companion galaxy.

\section{Implications}
\label{sec:implications}

Regardless of the nature of Cygnus A-2, it is a rare class of object, and the serendipitous detection of such an event so close to one of the most active nuclei in the nearby Universe suggests some sort of physical connection to its local environment.

If it is a rare type of supernova, the detection in this unusual location suggests that star-formation in the Cygnus A inner environment proceeds in an unusual way that is particularly conducive to producing radio-luminous supernova explosions.  This could be due to an altered IMF with much larger numbers of ultra-massive stars, a larger fraction of massive stars in close binaries, an increased rate of rare explosions at very high metallicities, or some combination of these effects.  If so, shallow radio supernova surveys of other extreme galaxy environments may be a fruitful means of finding other such transients.    

If it is a tidal disruption event, this would confirm the inference of a hugely elevated rate ($\sim10^{-1}$ yr$^{-1}$) within ULIRG-like galaxies as suggested by \cite{Tadhunter+2017}.

The most likely possibility, in our view, is that it represents an outburst from an active galactic nucleus due to a rapid increase in the accretion rate.  The inferred order-of-magnitude increase in flux over a ten-year timespan is also somewhat unusual among blindly-selected quasars, and may indicate that the configuration of this system (a secondary, lower-mass supermassive black hole in a distant orbit around a larger one) is particularly conducive to this sort of variability.

Perhaps the most interesting implication of this model, however, is the possibility of a connection between A-2 and the Cygnus A primary with its powerful jet.  Assuming a true offset close to the projected 460 pc and a mass interior to this of $\sim10^{10}\,{\rm M}_\odot$, the orbital timescale for the black hole is approximately $10^7$ yr.  This is quite similar to the timescale over which the Cygnus A jet system has been active \citep{Alexander+1984,Carilli+1991,Kino+2005}, a coincidence also noted by \cite{Canalizo+2003}.  This may be due to chance, but it is conceivable that the connection is a direct one, with the secondary black hole (and its surrounding halo) playing a key role in triggering or regulating the inflow associated with the current jet episode via its gravitational influence \citep{Hernquist+1989}.

Over a longer timescale, A-2 (and its former host galaxy) may also have been responsible for setting up the conditions necessary for the Cygnus A jet and nuclear starburst to form in the first place.  Quasars and radio galaxies are widely believed to result from mergers, and while the presence of ongoing major mergers is often obvious from the morphology of the host galaxies of these objects, for many examples (including Cygnus A itself) the host provides no direct clue to the nature or even presence of a merging companion.  If the active nucleus was at the center of the galaxy which delivered the gas to the center of Cygnus A's elliptical host, further observations may be able to shed light on the age, mass, and nature of the merging galaxy responsible.  

Our discovery may also help understand binary black hole evolution in a broader sense.  Only a few sub-kpc binaries are known \citep{Max+2005,Comerford+2015}, despite the important implications of these systems ranging from the nanoHz gravity wave background to black hole growth and AGN feeding.  The discovery that the most iconic powerful radio galaxy may be a binary SMBH argues that SMBH binarity may be more prevalent, and more important, than previously considered.  Future high-resolution, high-dynamic-range imaging of massive, luminous galaxy mergers similar to Cygnus A could lead to the discovery of additional examples of close, active binaries.  This would provide better constraints on the prevalence of this phenomenon and new insights into the SMBH inspiral process, its relation to the surrounding environment, and its implications for AGN triggering. \\[0.5cm]

\vskip 0.2cm

\acknowledgments

D.A.P.\ acknowledges past support from a Marie Sklodowska-Curie Individual Fellowship within the Horizon 2020 European Union (EU) Framework Programme for Research and Innovation (H2020-MSCA-IF-2014-660113).  

The National Radio Astronomy Observatory and the Long Baseline Observatory are facilites of the National Science Foundation operated under cooperative agreement by Associated Universities, Inc.  We acknowledge the use of imaging acquired from the Gemini Observatory Archive.  Gemini Observatory is operated by the Association of Universities for Research in Astronomy, Inc., under a cooperative agreement with the NSF on behalf of the Gemini partnership.

We thank T. Krichbaum and C. Max for providing us with their reduced archival observations.  We also thank the anonymous referee for providing useful input and suggestions that improved the quality of the paper.  We acknowledge useful comments and feedback from J.~Comerford, K.~Nyland, D.~A.~Kann, M.~Vestergaard, and A.~V.~Filippenko.


\bibliographystyle{apj}

\end{document}